\documentclass[12pt]{article}

\advance \textheight by \topskip
\makeatletter
\@addtoreset{equation}{section}
 
\makeatother

\def\bz{\bar z}
\def\R{\mathbb R}
\def\C{\mathbb C}
\def\N{\mathbb N}
\def\Z{\mathbb Z}

\def\D{\mathcal D}
\def\t{\mathfrak t}
\def\I{\hbox{{1\hskip -5.8pt 1}\hskip -3.35pt I}}
\usepackage{amsmath,amssymb}
\begin{document}
\title{
%\begin{flushright}
%{\small USACH-FM-02/02\\[-3mm]
%hep-th/0202077}\\[1.0cm]
%\end{flushright}
{\bf Nonlinear Holomorphic Supersymmetry on
Riemann Surfaces}}

\author{{\sf Sergey M. Klishevich${}^{a,b}$}\thanks{
E-mail: sklishev@lauca.usach.cl}
{\sf\ and Mikhail S. Plyushchay${}^{a,b}$}\thanks{
E-mail: mplyushc@lauca.usach.cl}
\\
{\small {\it ${}^a$Departamento de F\'{\i}sica,
Universidad de Santiago de Chile,
Casilla 307, Santiago 2, Chile}}\\
{\small {\it ${}^b$Institute for High Energy Physics,
Protvino, Russia}}}
\date{}

\maketitle

\vskip-1.0cm

\begin{abstract}
 We investigate the nonlinear holomorphic supersymmetry for
 quantum-mechanical systems on Riemann surfaces subjected to
 an external magnetic field. The realization is shown to be
 possible only for Riemann surfaces with constant curvature
 metrics. The cases of the sphere and Lobachevski plane are
 elaborated in detail. The partial algebraization of the
 spectrum of the corresponding Hamiltonians is proved by the
 reduction to one-dimensional quasi-exactly solvable
 $\mathfrak{sl}(2,\R)$ families. It is found that these
 families possess the ``duality'' transformations, which
 form a discrete group of symmetries of the corresponding 1D
 potentials and partially relate the spectra of different 2D
 systems. The algebraic structure of the systems on the
 sphere and hyperbolic plane is explored in the context of
 the Onsager algebra associated with the nonlinear
 holomorphic supersymmetry. Inspired by this analysis, a
 general algebraic method for obtaining the covariant form
 of integrals of motion of the quantum systems in external
 fields is proposed.
\end{abstract}

\newpage
\section{Introduction}

In the pioneer paper \cite{andrianov},
the usual linear supersymmetry \cite{cooper}
was generalized by employing
the higher-derivative supercharges.
The characteristic
property of such a generalization
is the polynomiality
of the corresponding superalgebra
in even integrals of motion of the system. This
makes the polynomial (nonlinear)
supersymmetry to be similar to the
Yangian and finite W-algebras \cite{yangian1,yangian2,Walg}.
The supersymmetry of such a type was found in
various physical models
\cite{ns1,ns2,ns3,susy-pb,susy-pf,ns4}.
This provides a solid background for physical
interest in the nonlinear supersymmetry.

The nonlinear holomorphic
supersymmetry ($n$-HSUSY)
\cite{d1,d2,kiev,nsusy}
is a natural generalization of the usual
linear \cite{cooper}
and higher-derivative (polynomial)
\cite{andrianov,susy-pb,susy-pf} supersymmetries.
Its construction was triggered by the
observation of the quantum anomaly
problem, which appears under attempt
to quantize the classical one-dimensional
system with nonlinear supersymmetry
of arbitrary order $n\in\N$, $n>1$
\cite{susy-pb}.
The important result in resolving this problem
was obtained in Ref. \cite{d1},
where we showed that
the anomaly-free quantization
is possible only for the peculiar
class of the superpotentials,
for which the corresponding
quantum $n$-HSUSY turns out to be
directly related to the quasi-exactly solvable
(QES) systems
\cite{turbiner,shifman,ushv,olv}.
Such one-dimensional quantum
mechanical systems were also
independently discussed
within the framework of the
so called Type A $\cal N$-fold
supersymmetry \cite{Nfold,Nfold1,aoyama,aoyama1,dorey},
where, in particular, the equivalence
between the nonlinear supersymmetry and
the $\mathfrak{sl}(2,\R)$ scheme for
the 1D QES systems was demonstrated \cite{aoyama}.

On the other hand, nowadays, the two-dimensional dynamics
attracts a considerable attention in the context of the
both, linear \cite{cooper,spin2,spin3,spin4}, and nonlinear
\cite{andrianov-2d,cannata,kuru}, supersymmetries. The
particularly interesting 2D system, related to the quantum
Hall effect \cite{iengo,li,hall}, is the non-relativistic
charged spin-1/2 particle in a stationary magnetic field
\cite{Jackiw,dunne,spin2}. In Ref. \cite{d2}, we generalized
the $n$-HSUSY to the 2D case represented by the system of a
charged spin-1/2 particle with gyromagnetic ratio $2n$ (or,
spin-$n/2$ particle with gyromagnetic ratio $2$) moving in
the plane with magnetic field of a specific form. The most
important result of such a generalization was the
observation of some universal nonlinear algebraic relations
underlying the $n$-HSUSY.

In the recent paper \cite{nsusy}, the nonlinear relations of
Ref. \cite{d2} were identified as the Dolan-Grady relations
\cite{dg}. As a result, the construction of the $n$-HSUSY
was reduced to the following universal
(representation-independent) algebraic structure. Its
supercharges have the form of (anti)holomorphic polynomials
in the two mutually conjugate operators (generating
elements), in whose terms all the components of the
$n$-HSUSY construction are built. It is these generating
elements that obey the Dolan-Grady relations and produce the
associated infinite-dimensional Onsager algebra
\cite{onsager}. The knowledge of the universal algebraic
construction essentially facilitated the search for the
central charges of the $n$-HSUSY, and simplified the
calculation of its nonlinear superalgebra. Having it, we
proposed a generalization of the $n$-HSUSY in the form of
nonlinear pseudo-supersymmetry \cite{nsusy}, which is
related to the $PT$-symmetric quantum mechanics
\cite{bb,znojil,phsym}.

The power of the algebraic formulation of the $n$-HSUSY
\cite{nsusy} is that it does not refer to the nature of the
space-time manifold which can be lattice, continuum or
noncommutative space. In this sense the $n$-HSUSY is
a direct algebraic generalization of the usual quantum-
mechanical supersymmetry \cite{cooper}.

The present paper is devoted to investigation of the
quantum mechanical problem of realization of the $n$-HSUSY
on Riemann surfaces with an inhomogeneous magnetic field.
Applying the algebraic construction of Ref. \cite{nsusy}, we
find a necessary condition for such a realization, and
discuss the spectral problem in the two possible
(non-trivial) cases of the sphere and Lobachevski plane
subjected to an external axial magnetic field of a peculiar
form. Reducing  the $n$-HSUSY systems to one
dimension, we find the new discrete
symmetry of the 1D QES potentials, which
induces the ``duality'' transformations between the
associated distinct $\mathfrak{sl}(2,\R)$ schemes.
The analysis of the axial symmetry of the 2D $n$-HSUSY
systems
in terms of the intrinsic algebra generated by the covariant
derivatives allows us to propose
(beyond the context of supersymmetry)
the general algebraic method
for
obtaining
the covariant form of integrals of motion of quantum
systems in a curved space in the presence of
an external
gauge field.

The paper is organized as follows. In Section 2 we discuss
the realization of the $n$-HSUSY on Riemann surfaces with a
magnetic field, and find the zero modes corresponding to the
cases of the sphere and Lobachevski plane. In Section 3 we
demonstrate that the spectral problem for these two cases
can be partially solved by a reduction to one dimension. The
obtained 1D systems admit distinct $\mathfrak{sl}(2,\R)$
representations, that is reflected in existence of the
discrete symmetry of the potentials. This symmetry is
investigated in Section 4. In Section 5 we study the
algebraic content of the systems on the sphere and
hyperbolic plane in the context of the associated
infinite-dimensional contracted Onsager algebra
\cite{nsusy}. Section 6 develops the
algebraic approach for finding
the covariant form of
the integrals of motion of the quantum systems in an
external
field with symmetry. In Section 7 we discuss the
obtained results and specify some problems to be interesting
for further consideration.

%%%%%%%%%%%%%%%%%%%%%%%%%%%%%
\section{$n$-HSUSY and Riemann surfaces}

\subsection{$n$-HSUSY algebraic structure}
The system with nonlinear holomorphic supersymmetry is
described by the Hamiltonian \cite{nsusy}
\begin{equation}\label{sH}
 \mathcal H_n=\frac 14\left\{\bar Z,\;Z\right\} +
 \frac n4\left[Z,\;\bar Z\right]\sigma_3.
\end{equation}
Here $\sigma_3$ is the diagonal Pauli
matrix, $n\in\N$, and
the mutually conjugate operators $Z$ and $\bar Z$ obey the
nonlinear
Dolan-Grady
relations
\begin{align}\label{dgr}
 \left[Z,\:\left[Z,\:\left[Z,\:\bar Z\right]\right]\right]&=
 \omega^2\left[Z,\:\bar Z\right],&
 \left[\bar Z,\:\left[\bar Z,\:
 \left[Z,\:\bar Z\right]\right]\right]&
 =\bar\omega^2\left[Z,\:\bar Z\right].
\end{align}
These relations
guarantee the existence of the odd integrals
of motion (supercharges),
$Q_n$ and $\bar{Q}_n$, which are defined
by the recurrent relations
\begin{align}
 Q_{n}&=\left(Z^2-\left(\tfrac{n-1}2\right)^2
 \omega^2\right)Q_{n-2},&
 Q_0&=\sigma_+,&Q_1&=Z\sigma_+,
 \notag\\[-2.8mm]\label{Qalg}\\[-2.8mm]\notag
 \bar Q_{n}&=\left(\bar Z^2-\left(\tfrac{n-1}2\right)^2
 \bar\omega^2\right)\bar Q_{n-2},&
 \bar Q_0&=\sigma_-,&\bar Q_1&=\bar Z\sigma_-,
\end{align}
containing a constant parameter
$\omega \in\C$,
$\bar{\omega}=\omega^*$,
and $\sigma_\pm=\frac{1}{2}(\sigma_1\pm i\sigma_2)$.
The supercharges $Q_n$, $\bar{Q}_n$
generate a nonlinear superalgebra
of order $n$ \cite{nsusy}.
In Section 5 we shall return to the discussion
of the algebraic construction underlying the
nonlinear holomorphic supersymmetry.

\subsection{Fixing geometry}
\label{geom}

We are going to investigate the problem of
realization of the $n$-HSUSY on a 2D surface with
a nontrivial Riemann geometry.
The metrics of any two-dimensional surface can be
represented {\it locally} in the conformally flat form,
\begin{equation}\label{m}
 ds^2=\rho(z,{\bar z})dz d\bar z=
 g_{z\bz}dzd\bz+g_{\bz z}d\bz dz,
\end{equation}
where $z=x^1+ix^2$, $\bz=x^1-ix^2$ are
the isothermal
coordinates.
Given the
isothermal coordinates on the neighbouring patches,
the metrics on the patches are related by conformal
transformations.
Hence, we have
\begin{align*}
 g_{z\bar z}&=\frac\rho 2,&
 g^{z\bar z}&=\frac 2\rho,&
 g_{zz}&=g_{\bz\bz}=0,&
 \sqrt g&=\rho,
\end{align*}
where $g=\mathop{\sf det}g_{\mu\nu}$.

Let us
choose the operators $Z$ and $\bar Z$ in the form of
covariant derivatives,
\begin{align}\label{Z}
 Z&=\nabla_z=D+\partial\log\rho\cdot\hat S, &
 \bar Z&=-\nabla^z=-g^{z\bar z}\bar D,
\end{align}
where the operator $\hat S$ defines the helicity, or the
conformal spin of a tensor
$T^{(s)}$, $\hat ST^{(s)}=sT^{(s)}$,
$s\in \mathbb Z$, while $D=\partial+i\hat eA$ and
$\bar D=\bar\partial+i\hat e\bar A$
are the $U(1)$ covariant derivatives with $A$, $\bar A$ to
be the components of the vector potential corresponding to
the case of stationary magnetic field. The operator $\hat e$
gives the electric charge of the tensor field.
By definition, the derivative $\nabla_z$ maps the
tensor fields
of the helicity $s$ into those of
the helicity $s-1$, while $\nabla^z$
transforms the tensors of the helicity $s$ into those of
the helicity $s+1$. The operators (\ref{Z})
are mutually conjugate with
respect to the scalar product \cite{nakahara}
\begin{equation}\label{scal}
 \langle T^{(s)},U^{(s)}\rangle=\int d^2z\sqrt g
 \left(g_{z\bar z}\right)^s\bar T^{(s)}U^{(s)}
\end{equation}
defined on the space of the tensor fields
satisfying the appropriate boundary
conditions.

In representation (\ref{Z}) we have
\begin{equation}
[Z,\bar Z]=B\hat e -
\frac{1}{2}\mathcal R \hat S,
\label{ZZBR}
\end{equation}
where
the scalar curvature of the Riemann surface
$\mathcal R$ and the magnetic field $B$ are
\begin{equation}
\mathcal R=-\frac 4\rho\partial\bar\partial\ln\rho,\qquad
B\hat e =-\frac{2}{\rho}[D,\bar D].
\label{BR}
\end{equation}
As a result,
the nonlinear Dolan-Grady relations (\ref{dgr})
take the form
\begin{align}
 \left((\nabla_z^2 B) -\omega^2B\right)\hat e
 - \frac 12\left((\nabla_z^2\mathcal R)
 -\omega^2\mathcal R\right)\hat S
 - (\nabla_z\mathcal R)\nabla_z&=0,
 \notag\\[-3mm]\label{rc}\\[-3mm]\notag
 \left((\nabla^z{}^2B)-\bar\omega^2B\right)\hat e
 - \frac 12\left((\nabla^z{}^2\mathcal R)
 -\bar\omega^2\mathcal R\right)\hat S
 + (\nabla^z\mathcal R)\nabla^z&=0.
\end{align}
The relations (\ref{rc}) have to be satisfied as
the operator identities.
Hence, the coefficients at the operators $\hat e$,
$\hat S$,
$\nabla_z$ and $\nabla^z$
have to vanish independently.
The coefficients at the
covariant derivatives vanish only
when the scalar curvature
is a constant,
\begin{equation}\label{r=c}
 \mathcal R=const.
\end{equation}
With this condition, the
coefficients at the operator $\hat S$
result in the equation
$\omega^2\mathcal R=0$, and, hence,
there are two possibilities: either $\omega=0$, or
$\mathcal R=0$. The latter case corresponds to
the plane geometry, and it
was analysed in detail in Ref. \cite{d2}.
Therefore, in what follows we shall
discuss the former case.

For $\omega=0$, there arise the
contracted Dolan-Grady relations,
\begin{align}\label{cdgr}
 \left[Z,\:\left[Z,\:\left[Z,\:
 \bar Z\right]\right]\right]&=0,&
 \left[\bar Z,\:\left[\bar Z,\:
 \left[Z,\:\bar Z\right]\right]\right]&=0.
\end{align}
The nonlinear supersymmetry associated with the
contracted Dolan-Grady relations (\ref{cdgr}) is generated
by the supercharges of the form
\begin{equation}\label{sQ}
 Q_n=Z^n\sigma_+,\hskip 2cm
 \bar Q_n=\bar Z^n\sigma_-,
\end{equation}
corresponding to the limit
$\omega\to 0$
of the general case (\ref{Qalg}).

One notes that, in principle,
the restriction (\ref{r=c}) can be
overcome if to endow the Riemann structure of the 2D
surface with the torsion. Then,
for $\omega\neq 0$, the Dolan-Grady relations
(\ref{dgr}) will result in a set of nonlinear equations
on the scalar curvature and on the torsion.
However, here we restrict
ourselves by the torsion-free case with
the relation (\ref{r=c}) to be the
necessary condition for existence of
the $n$-HSUSY on a Riemann surface.

The general uniformization theorem \cite{rs} implies
that, up to the holomorphic equivalence,
there are just three distinct simply connected Riemann
surfaces: a) the sphere $S^2$
with the conformal factor
\begin{equation}\label{S2}
 \rho(z,{\bar z})=\frac 1{(1+\beta^2z{\bar z})^2},
\end{equation}
where $z\in\C\cup\{\infty\}$; b) the complex plane $\C$ with
the flat metric, $\rho=1$; c) the Lobachevski (hyperbolic)
plane
with the metric (\ref{m}) defined on
the disk
$|z|<\beta^{-1}$
by
\begin{equation}\label{H}
 \rho(z,{\bar z})=\frac 1{(1-\beta^2z{\bar z})^2}.
\end{equation}
In Eqs. (\ref{S2}),
(\ref{H}),
the positive parameter $\beta$ defines
the value of the scalar curvature,
$\mathcal R=\pm 8\beta^2$,
where the plus corresponds to the
sphere, while the minus does to the hyperbolic plane. In the
case of the sphere the parameter $\beta$ is related to its
radius as $\beta^{-1}=2R$.

In what follows we will mainly analyse
the case of the sphere,
while all the corresponding results for
the Lobachevski plane
can formally be
obtained by the
change $\beta\to i\beta$.
At the same time,
the construction of the $n$-HSUSY
on the plane \cite{d2} can be
reproduced
from the cases corresponding to
the surfaces
(\ref{S2}) and (\ref{H}) via
the appropriate limit procedure
$\beta\rightarrow 0$.

\subsection{Fixing magnetic field}
We have not analysed yet the conditions
corresponding to the vanishing of the
coefficients at
$\hat e$.
They produce the equations on
the
external magnetic field.
In correspondence with Eq. (\ref{BR}),
the
magnetic field is given locally by the relation
\begin{equation*}%\label{defB}
 Bdv=d\mathcal A,
\end{equation*}
where $dv=\left(i\sqrt{g}/2\right)dz\wedge d\bz$ is the
area (volume) element, and
${\cal A}=Adz+\bar Ad\bar z$.
In correspondence with (\ref{rc}), for $\omega=0$
we have the following equations on the magnetic field
$B=-ig^{z\bar z}\left(\partial\bar A-\bar\partial A\right)$:
\begin{align*}%\label{eB}
 \partial\left(\frac 1{\rho(z,\bar z)}
 \partial B(z,\bar z)\right)&=0, &
 \bar\partial\left(\frac 1{\rho(z,\bar z)}
 \bar\partial B(z,\bar z)\right)&=0.
\end{align*}
For the both cases (\ref{S2})
and (\ref{H}), their solution can be
represented in the form
\begin{equation}\label{B}
 B(z,\bar z)=c_1\sqrt{\rho(z,\bar z)}+c_0,
 \end{equation}
with $c_0,c_1\in\R$.
In the case of the sphere the
(normalized for $2\pi$)
total magnetic flux,
\begin{align}\label{flux}
 2\pi \Phi=\int\limits_{S^2}Bdv=
 2\pi \int\limits_0^\infty B(r)\rho(r)rdr=
 \frac{\pi(2c_0+c_1)}{2\beta^2},
\end{align}
is quantized, $\Phi\in\mathbb Z$. Here we have
used the polar
representation of the complex coordinates,
$z=re^{i\varphi}$, $\bz=re^{-i\varphi}$.

In the case of the hyperbolic plane,
the magnetic flux is divergent,
but the relation (\ref{flux})
(with the substitution
$\beta^2\rightarrow -\beta^2$)
still can be used
in the form $\Phi=-(2c_0+c_1)/(4\beta^2)$
as a mere combination of the parameters
which can acquire any real value.

In what follows, we fix the charge of the system
to be $e=+1$.

\subsection{Zero modes}
Let us find the zero mode subspace
of the $n$-HSUSY realized on the
Riemann surfaces (\ref{S2}), (\ref{H})
with magnetic field (\ref{B}).
To this end we note that
the space of the states of the system
(\ref{sH}) is a direct
sum of the
two Hilbert spaces,
$\mathcal F_-\oplus\mathcal F_+$,
formed by the states
$\Psi^-\in \mathcal F_-$
and
$\Psi^+\in \mathcal F_+$,
being the eigenstates of $\sigma_3$,
$\sigma_3\Psi^\mp=\mp \Psi^\mp$.
We will refer to the
spaces $\mathcal F_-$ and $\mathcal F_+$,
respectively,
as to the ``
bosonic'' and ``fermionic'' subspaces.
We also assume that the ``bosonic'' wave functions are
the tensors of helicity $s$. Then, in accordance with the
structure of the supercharges (\ref{sQ}),
the ``fermionic'' wave functions will
be the tensors of the helicity $s-n$.

The Hamiltonian acting on the ``bosonic'' wave functions can
be rewritten in the complex coordinates as
\begin{equation}\label{zH}
 H_n^{(s)}={}-\frac 1{2\rho}\left(\left\{\bar D,D\right\}
 +2s\partial\log\rho\cdot \bar D\right)
 -\frac n4B-{s(n+1)}\beta^2.
\end{equation}
The upper index in the parentheses denotes the helicity of
the tensor that the Hamiltonian acts on. In these notations
the ``fermionic'' Hamiltonian is $H_{-n}^{(s-n)}$.

In the gauge
\begin{align}\label{gz}
 A&=-i\frac\bz8\left(c_1\rho+4\beta^2\Phi\sqrt\rho\right),&
 \bar A&=i\frac z8
 \left(c_1\rho + 4\beta^2\Phi\sqrt\rho\right),
\end{align}
corresponding to the magnetic field (\ref{B}),
(\ref{flux}), the states of the form
\begin{align}\label{zm-}
 \psi^{(-)}_n(z,\bz)&=
 \rho^{\frac\Phi 4-s+\frac{n-1}2}
 e^{\frac{c_1}{8\beta^2}\sqrt\rho}
 \sum_{k=0}^{n-1}\varphi^{(-)}_k(\bz)z^k,
 \\\notag
 \psi^{(+)}_n(z,\bz)&=
 \rho^{\frac{n-1}2-\frac\Phi 4}
 e^{-\frac{c_1}{8\beta^2}\sqrt\rho}
 \sum_{k=0}^{n-1}\varphi^{(+)}_k(z)\bz^k
\end{align}
are the zero modes of the supercharges $Q_n$ and $\bar Q_n$,
respectively.
In the case of the sphere,
the condition of normalizability fixes the form
of the functions $\varphi^{(\pm)}_k$,
\begin{align}\label{f-}
 \varphi^{(-)}_k(\bz)&=\sum_{q=0}^{2(n-s_\Phi-1)-k}
 \varphi^{(-)}_{kq}\bz^q,&%\\\notag
 \varphi^{(+)}_k(\bz)&=\sum_{q=0}^{2(s_\Phi-1)-k}
 \varphi^{(+)}_{kq}\bz^q,
\end{align}
where we have introduced the notation
\begin{equation}
s_\Phi=s-\frac{1}{2}\Phi.
\label{sphi}
\end{equation}
The physical sense of the parameter $s_\Phi$
will be found below under analysis of the contracted
Onsager algebra associated with the system.
The corresponding
number of the normalizable zero modes in the
``bosonic'' and ``fermionic'' sectors is given by
\begin{align*}%\label{Nbf}
 N_B&=\frac 12n\left(3n-4s_\Phi-1\right),&
 N_F&={}-\frac 12n\left(n-4s_\Phi+1\right).
\end{align*}
Here we assume that $N_B,N_F\in\mathbb N$;
for other values of the parameters
$n$, $s$ and $\Phi$ the corresponding
number of zero modes is implied to be equal to zero.
One can verify that in the spheric case $N_B+N_F>0$ for any
$n\in\N$, and, therefore, the nonlinear supersymmetry of the
corresponding system is always unbroken.

In the case of the hyperbolic plane, the normalizability is
managed by the sign of the parameter $c_1$, and the
corresponding space of zero modes is infinite-dimensional.
Moreover on the hyperbolic plane, unlike the sphere case,
the supercharges $Q_n$ and $\bar Q_n$ can not have zero
modes simultaneously.

So, the nonlinear holomorphic supersymmetry of
quantum-mechanical systems on 2D surfaces leads to the
strong restrictions on configurations of the external
geometry and magnetic field. Moreover, the systems with
the $n$-HSUSY on the sphere are quasi-exactly solvable since
the
space of zero modes of the supercharges is
finite-dimensional in this case. In the next section we will
demonstrate that the same is also true for the systems on
the hyperplane.

%%%%%%%%%%%%%%%%%%%%%%%%%%%%%%%%%%%%%%%
\section{Reduction to $\mathfrak{sl}(2,\R)$ quasi-exactly
solvable families}
\label{sl2}

In this section we consider the spectral problem for the
``bosonic'' Hamiltonian (\ref{zH}) and show that it can be
reduced to a one-dimensional problem corresponding to
$\mathfrak{sl}(2,\R)$ quasi-exactly solvable families of
the systems.
The spectral problem in the ``fermionic''
sector can be analysed in a similar way.

{}First, we note that the functions of the form
$z^mf(\bz z)$, $m\in\Z$, constitute a subspace
invariant with respect
to the action of the Hamiltonian (\ref{zH}) in the gauge
(\ref{gz}). The parameter $m$, as will be shown,
is the quantum number
corresponding to the integral (\ref{Wc})
associated with the axial symmetry of the system.
Departing
from the form of zero modes (\ref{zm-}),
we
search for the eigenfunctions of the form
\begin{equation}\label{anz}
 \psi(z,\bz)=z^m(\bz z)^a(\rho(\bz z))^b
 e^{c\sqrt{\rho(\bz z)}}\varphi(\bz z).
\end{equation}
Now, we introduce the variable
$$
 y=\sqrt{\rho(\bz z)}
$$
with the domain $0\leq y\leq 1$.
The anzatz (\ref{anz})
reduces the two-dimensional Hamiltonian (\ref{zH})
to the one-dimensional operator
\begin{equation}\label{Hqes}
 H_{sl(2,\R)}=2\beta^2\left(T_0^2 - T_-T_0\right)
 +\alpha_+ T_+ + \alpha_0T_0+\alpha_-T_- + const
\end{equation}
if the parameters $a$, $b$ and $c$
are chosen in the following way:
\begin{align}\label{abc}
 a&=\frac{\sigma-1}2m,&
 b&=\frac{\nu}2s_\Phi-\frac{s}{2}+\frac{\sigma-\nu}4m,&
 c&=\mu\frac{c_1}{8\beta^2}.
\end{align}
The parameters
$\mu$, $\nu$ and
$\sigma$ acquire the values $\pm 1$,
while the operators
\begin{align}
\label{sl2rg}
 T_+&=(N-1)y-y^2\frac d{dy},&
 T_0&=y\frac d{dy}-\frac{N-1}2,&
 T_-&=\frac d{dy}
\end{align}
generate the
$\mathfrak{sl}(2,\R)$
algebra with the Casimir operator
fixed by the parameter $N$.
When $N\in\N$, the
realization
(\ref{sl2rg})
gives an irreducible $N$-dimensional
representation
on the space of polynomials of the
degree up to $N-1$.
In this case the operator (\ref{Hqes})
corresponds to some QES system
\cite{turbiner,shifman,ushv}.

The corresponding parameters of the $\mathfrak{sl}(2,\R)$
scheme, $N$, $\alpha_\pm$ and $\alpha_0$, read as
\begin{align}
 N&=\mu n + \frac{\nu-\sigma}2m
 - (\mu+\nu)s_\Phi,\notag\\\notag
 \alpha_+&=-\mu\frac{c_1}2,\\[-3mm]\label{N}\\[-3mm]\notag
 \alpha_0&={}\beta^2\bigl(2\mu n
 - \left(\nu-\sigma\right)m
 - \left(\mu-\nu\right)2s_\Phi\bigr)
 - \mu\frac{c_1}2,\\\notag
 \alpha_-&=\beta^2\left(1-\mu n +\frac{3\nu+\sigma}2m
 + (\mu-3\nu)s_\Phi
 \right).
\end{align}
So, we have formally $2^3=8$ solutions parametrised by the
triple
$(\mu, \nu,\sigma)$. Of course, the solutions (\ref{N})
realize a true quasi-exactly solvable system only if the
condition $N\in\N$ is satisfied. For example, the solution
with the triple $(-1,+1,+1)$ does not correspond to any QES
system since in this case $N=-n$, while throughout the paper
we assume $n\in\N$. But as it will be shown further,
this solution together with the other solutions reflects
some discrete symmetry of the resulting one-dimensional
system.
We also note that
under the change $(\mu,\nu,\sigma)\to
(-\mu,-\nu,-\sigma)$ in (\ref{N}),
the parameter $N$ changes its sign.
This means that for any
choice of the parameters $n$, $m$
and $s_\Phi$,
at most four solutions from
(\ref{N}) can realize the proper finite-dimensional
$\mathfrak{sl}(2,\R)$ representations of the QES system.

{}From the 2D viewpoint, the conditions of regularity at
zero
and of finiteness at infinity\footnote{Since we consider the
sphere, it is necessary to require the regularity of all the
wave functions for $|z|\to\infty$.} for the functions
(\ref{anz}),
(\ref{abc}) lead,
correspondingly,
to the constraints
\begin{equation}\label{reg}
 \sigma m\ge 0,\hskip 2cm
 \nu m\le 2(\nu s_\Phi-s).
\end{equation}
The condition of normalizability of
the wave functions results, in turn, in the inequality
\begin{equation}\label{norm}
 \nu\left(2s_\Phi-m\right)+1>0.
\end{equation}
In all the relations (\ref{reg}), (\ref{norm}) we
imply that the parameters $m$, $s$ and $\Phi$ accept integer
values only.

Changing the variable,
$$
 y=\cos^2\beta x,
$$
with $0\leq x\leq \frac\pi{2\beta}$,
and following the standard procedure
of the $\mathfrak{sl}(2,\R)$ scheme for QES systems
\cite{turbiner,shifman,ushv}, one
can transform the operator (\ref{Hqes}) to the
one-dimensional Hamiltonian of the standard form,
$$
 H={}-\frac 12\frac{d^2}{dx^2} + V(x).
$$
All the solutions (\ref{N}) lead to the very potential
\begin{align}
 V(x)&={}-\frac{c_1^2}{32\beta^2}{\cos^4\beta x} +
 \frac{c_1}4\left(2\left(s_\Phi-n\right)
 +\frac{c_1}{8\beta^2}\right)\cos^2\beta x
 \notag\\[-3mm]\label{V}\\[-3mm]\notag&{}
 +\frac{\beta^2}8\left(4m^2-1\right)\cot^2\beta x
 + \frac{\beta^2}8\left(4\left(2s_\Phi-m\right)^2
 -1\right)\tan^2\beta x.
\end{align}
This means that the potential (\ref{V})
possesses some discrete
symmetry, which relates the solutions (\ref{N}). We
discuss this symmetry in the next section.

The potential (\ref{V}) has the following behaviour near the
point $x=0$:
$$
 V(x)=const+\frac{4m^2-1}{8x^2}+\mathcal O(x^2).
$$
Therefore, in the case $m=\pm\frac 12$, it is formally
possible to extend the domain of definition of $x$ to the
symmetric interval
$-\frac\pi{2\beta}\leq x\leq \frac\pi{2\beta}$ (or
to $\R$ in the case of the hyperbolic plane, see below).

The case of the plane corresponds to the limit $\beta\to 0$.
Rescaling and shifting the parameters,
$$
 c_0\to c_0+\frac{c_1}{\beta^2}, \hskip 2cm
 c_1\to-\frac{c_1}{\beta^2},
$$
in this limit
we recover from Eq. (\ref{B})
the
magnetic field of the form \cite{d2}
\begin{equation*}%\label{Bplane}
 B(r)=c_1r^2+c_0,\hskip 1.5cm
 r^2=\bar{z}z,
\end{equation*}
and find that the described reduction
results in the one-dimensional QES potential
\begin{equation}\label{x6}
 V(x)=\frac{c_1^2}{32}x^6 + \frac{c_1c_0}8x^4
 + \frac 18\left(c_0^2-2c_1(2n - m)\right)x^2
 + \frac{4m^2-1}{8x^2} + const,
\end{equation}
defined on the half-line $0\le x<\infty$.
Hence, the proper limit procedure reproduces
correctly the results derived earlier in Ref. \cite{d2}.

The formulas corresponding
to the case of the hyperbolic
plane are obtained by the formal substitution
$\beta\to i\beta$ with the change of the domains of
the variables $y$ and $x$ to $1\leq y<\infty$ and
$0\leq x<\infty$. Under such transformations, the
trigonometric potential (\ref{V}) is
converted into the hyperbolic QES potential
\begin{align}
 V(x)&=\frac{c_1^2}{32\beta^2}{\cosh^4\beta x} +
 \frac{c_1}4\left(2\left(s_\Phi-n\right)
 -\frac{c_1}{8\beta^2}\right)\cosh^2\beta x
 \notag\\[-3mm]\label{Vh}\\[-3mm]\notag&{}
 +\frac{\beta^2}8\left(4m^2-1\right)\coth^2\beta x
 + \frac{\beta^2}8\left(4\left(2s_\Phi-m\right)^2
 -1\right)\tanh^2\beta x.
\end{align}
The condition of regularity at zero is the same as for the
sphere, $\sigma m\ge 0$, while the condition of finiteness
at
infinity has to be discarded. Besides, the normalizability
of the functions (\ref{anz}), (\ref{abc}) is changed for
$\mu c_1>0$ instead of (\ref{norm}). Since the magnetic flux
diverges in this case, the parameter combination
$\Phi$, specified at the very end of Section 2.3,
can acquire any real value. Therefore, in general, only
the solutions $(+1,-1,+1)$, $(+1,-1,-1)$ and $(-1,+1,-1)$
can produce the proper finite-dimensional
$\mathfrak{sl}(2,\R)$ representations, while other values of
the triple $(\mu,\nu,\sigma)$ serve only for integer values
of the parameter $\Phi$.

%%%%%%%%%%%%%%%%%%%
\section{``Duality'' transformations}

In the previous section we have seen that
the reduction of the two-dimensional
Hamiltonian, associated with the $n$-HSUSY
on the sphere or hyperbolic plane
with magnetic field (\ref{B}),
results in the one-dimensional quantum system
with the potential (\ref{V}) or (\ref{Vh}), respectively.
The resulting 1D Hamiltonians can be transformed
to the QES operator (\ref{Hqes}) with appearance
of the eight different
families of the $\mathfrak{sl}(2,\R)$ schemes.
Here we investigate the question what these different
families
of the $\mathfrak{sl}(2,\R)$ schemes
mean from the point of view of the
corresponding one-dimensional quantum mechanical
QES system.

First, we note that the potential (\ref{V}) is
effectively
four-parametric since
the parameter $\beta$ can be absorbed
by rescaling
$x\rightarrow \beta^{-1} x$,
$c_1\rightarrow \beta^2 c_1$,
$H\rightarrow \beta^{-2}H$,
while $s$ and $\Phi$ enter only in the combination
(\ref{sphi}).
The parameters $n$, $2s_\Phi$
and $m$ take integer values,
and $c_1$ is real. These parameters enter into the potential
(\ref{V}) in quadratic combinations. This motivates us to
search for the linear transformations of the parameters,
which leave the potential to be invariant. It is convenient
to represent them in the matrix form
\begin{equation}\label{trans}
 Y'=GY,
\end{equation}
with $Y^T=(n,2s_\Phi,m,c_1)$.
Let us introduce the following matrices:
\begin{align}\label{G}
 G_1&=\begin{pmatrix}
 -1&1& 0&0\\
 0&1& 0&0\\
 0& 0&1&0\\
 0& 0& 0&-1\\
 \end{pmatrix},&
 G_2&=\begin{pmatrix}
 1&-1&1&0\\
 0&-1&2&0\\
 0& 0&1&0\\
 0& 0&0&1\\
 \end{pmatrix},&
 G_3&=\begin{pmatrix}
 1&0&-1&0\\
 0&1&-2&0\\
 0&0&-1&0\\
 0&0& 0&1\\
 \end{pmatrix}.
\end{align}
One can verify that the potential (\ref{V}) stands invariant
under the transformations (\ref{trans}) generated by the set
of the mutually commuting matrices
\begin{equation}\label{Gmat}
 G=\I,\ -\I,\ G_i,\ -G_i,\qquad i=1,2,3,
\end{equation}
satisfying the relations
\begin{equation}\label{duali}
 G_1G_2G_3=-\I, \hskip 2cm G_1^2=G^2_2=G^2_3=\I.
\end{equation}
Here, the eight transformations produced by the
matrices $G$ on the space of the parameters $Y$
correspond to eight transformations
in terms of the parameters $(\mu,\nu,\sigma)$,
which connect different solutions (\ref{N})
at the level of the $\mathfrak{sl}(2,\R)$ schemes.
In particular, the transformations generated by
the $G_1$, $G_2$ and $G_3$ correspond,
respectively,
to the transformations
$(\mu,\nu,\sigma)\rightarrow (-\mu,\nu,\sigma)$,
$(\mu,-\nu,\sigma)$ and $(\mu,\nu,-\sigma)$.
{}From this point of view, the transformations
(\ref{trans})--(\ref{Gmat})
give some irreducible representation
of the Abelian discrete group of reflections in three axes.

Another nontrivial example of such discrete transformations
is given by the QES potential (\ref{x6})
related to the $\mathfrak{sl}(2,\R)$-operator
\begin{align}
\label{hx6}
 H_{QES}&={}-2T_0T_- + \beta_+T_+ + \beta_0T_0 + \beta_-T_-,
\end{align}
where the $\mathfrak{sl}(2,\R)$-generators
are given by Eq. (\ref{sl2rg})
with $y=x^2$ \cite{turbiner,turb}.
The complete set of
the discrete transformations (\ref{trans}) with
$Y^T=(n,m,c_0,c_1)$ is given in this case
by the matrices
$\I$, $-\I$, $\tilde{G}$, $-\tilde{G}$,
where
$$
 \tilde G=\begin{pmatrix}
 1&-1& 0&0\\
 0&-1& 0&0\\
 0& 0&1&0\\
 0& 0& 0&1
 \end{pmatrix}.
$$
In comparison with (\ref{V}), the case (\ref{x6}) with
$n,m\in\Z$, $c_0,c_1\in\R$, is more simple, and, hence,
more illustrative. Let us discuss here the
corresponding eigenfunctions, which can be found
algebraically. One can represent them formally
as
\begin{align}\label{psi}
 \psi(x)&=P_{N-1}\left(x^2\right)x^{\frac 12+\nu m}
 \exp\left(-\frac\mu{16}c_1x^4-\frac\mu 8c_0x^2\right),&
 \text{with} &&N&=\mu n-\frac{\mu+\nu}2m,
\end{align}
and $\mu$, $\nu$ acquiring the values $\pm 1$.
For the functions (\ref{psi}) to be normalized, one has to
impose some restrictions on the parameters. There are four
distinct cases, which can be marked by the pair $(\mu,\nu)$:
\begin{equation*}%\label{cases}
\begin{array}{rclll}
 (+1,-1)&:&N= n\in\N,&c_1>0,&m\in\Z_-;\\
 (+1,+1)&:&N=n-m\in\N,&c_1>0,&m\in\Z_+;\\
 (-1,+1)&:&N= -n\in\N,&c_1<0,&m\in\Z_+;\\
 (-1,-1)&:&N=m-n\in\N,&c_1<0,&m\in\Z_-.
\end{array}
\end{equation*}
The restrictions on $c_1$ arise from the requirement of the
normalizability, while those on $m$ do from the requirement
of the vanishing of the functions at $x=0$, since we
consider
the spectral problem on the half-line, $0\le x<\infty$. In
fact, the cases $(+1,-1)$, $(-1,+1)$ correspond to the
conventional $\mathfrak{sl}(2,\R)$ representation of the
potential (\ref{x6}) (see, e.g., Refs. \cite{shifman,turb}),
while those $(+1,+1)$, $(-1,-1)$ were not discussed earlier.
All the four cases are related by the discrete
transformations as it is reflected on the diagram:
$$
 \begin{array}{ccc}
 (+1,-1)&\stackrel{\displaystyle\tilde G}{
 \longleftarrow\hskip -1mm\longrightarrow}&
 (+1,+1)\\
 \uparrow&&\uparrow\\[-2mm]
 |\lefteqn{-\I}&&|\lefteqn{-\I}\\[-2mm]
 \downarrow&&\downarrow\\[-2mm]
 (-1,+1)&\stackrel{\displaystyle\tilde G}{
 \longleftarrow\hskip -1mm\longrightarrow}&(-1,-1)\\
 \end{array}
$$

In principle, one can treat the system
(\ref{x6}) from the
more general point of view by assuming that all the
parameters $n$, $m$, $c_0$ and $c_1$ are real.
In this case the Hamiltonian
can also be reduced to the form
(\ref{hx6}), but if there are no additional
restrictions on the parameters,
the $\mathfrak{sl}(2,\R)$
generators realize an infinite-dimensional representation of
the algebra. As a consequence, the 1D system with the
potential (\ref{x6}) is not quasi-exactly solvable.
Now, if any of the
restrictions $\pm n\in\N$ or $\pm(n-m)\in\N$ is imposed,
the 1D system under consideration becomes to be
quasi-exactly solvable since with such a restriction
the corresponding
representation of the $\mathfrak{sl}(2,\R)$
is finite-dimensional.
For example, when $m=\frac
12$,
the domain of definition of the potential (\ref{x6}) can be
extended to the whole real axis. Let $c_1>0$,
then the pair $(+1,-1)$ with $N=n\in\N$ gives
even eigenfunctions (\ref{psi}),
while the pair
$(+1,+1)$ with
$N=n-\frac 12\in\N$ provides
the odd eigenfunctions. These two
cases correspond to the two distinct
forms of the QES $x^6$-potential~\cite{shifman}.

Having the two examples, one can suppose that the existence
of the discrete symmetry transformations is general for all
the
one-dimensional QES $\mathfrak{sl}(2,\R)$ systems. Let us
give some simple arguments in favour of this conjecture.

Changing appropriately the
variable and realizing the
similarity
transformation,
one can represent any QES $\mathfrak{sl}(2,\R)$ Hamiltonian
in the
canonical form
\cite{turbiner,shifman,ushv,olv}
\begin{equation}\label{Hqesg}
 H_{QES}=-P_4(y)\frac{d^2}{dy^2}
 +\left(\frac{N-1}2P_4'(y)-P_2(y)\right)\frac{d}{dy}
 +\frac N2\left(P_2'(y)-\frac{N-1}6P_4''(y)\right),
\end{equation}
where $P_k(y)$ is a polynomial of the
$k$th degree. The
potential of the corresponding Schr\"odinger equation is
given by
\begin{equation}\label{Vg}
 V(x)=\frac{N(N+2)}{12}\left(\frac 34\frac{P_4'{}^2}{P_4}
 -P_4''\right)+\frac{N+1}4\left(2P_2'-P_2\frac{P_4'}{P_4}
 \right)+\frac 14\frac{P_2^2}{P_4}+const,
\end{equation}
where the right hand side is evaluated at $y=f^{-1}(x)$ with
$f^{-1}$ being the inverse function of elliptic integral
$$
 f(y)=\int\frac{dy}{\sqrt{P_4(y)}}.
$$
The canonical form (\ref{Hqesg}) is not unique because of
the
existence of a ``residual'' symmetry, which allows us to fix
essentially
the polynomial $P_4$ \cite{olv}. Therefore,
one can consider that coefficients of the polynomial $P_2$
and the discrete parameter $N$ span the whole parametric
space of the system. From (\ref{Vg}) it follows that the
parameters enter into the potential in quadratic
combinations (see the explicit set of QES potentials, e.g.,
in Ref. \cite{turb}). Therefore, appealing to the
cases
(\ref{V}) and (\ref{x6}), it seems plausible that the
transformations of
the
form (\ref{trans}) should also exist in the general case
(\ref{Vg}). Of course, in general, for the transformed
parameter $N$ to remain integer, one has to treat some
parameters as to be also integer.

Let us  stress that
if one treats a 1D QES system in
the framework of the corresponding 2D supersymmetric system,
then the discrete transformations can intertwine the
distinct sectors of the spectrum of the same 2D system, or
the
spectra of different 2D systems. This property
as well as the relation $G^2=1$ allows us to treat
(\ref{trans}) as some kind of ``duality''
transformations. For example, one can verify that the zero
modes (\ref{zm-}), (\ref{f-}) are spanned by the solutions
(\ref{abc}) with the triples
$(+1,-1,+1)$ and $(+1,-1,-1)$,
which are related by the ``duality'' transformation
generated by the $G_3$ from Eq. (\ref{G}).

In conclusion of this section we note that
in the context of QES
systems, some
similar duality transformations were discussed in Refs.
\cite{dual1,dual2,dual3}.
However,  the ``duality'' transformations
discussed here are essentially different from
the transformations of
Refs. \cite{dual1,dual2,dual3}.
In the latter case
the duality connects {\it different}
1D QES systems, while in the
present case it relates different parts of
the spectrum of one and the same 1D QES system.
These parts correspond to different
$\mathfrak{sl}(2,\R)$ schemes
of the given potential.

%%%%%%%%%%%%%%%%
\section{The contracted Onsager algebra}

In this section, following Ref. \cite{nsusy},
we analyse the algebraic structure
underlying the 2D spherical and hyperbolic
systems with the nonlinear holomorphic supersymmetry.

The operators $Z_0\equiv Z$ and $\bar Z_0\equiv \bar Z$
together with the contracted Dolan-Grady relations
(\ref{cdgr}) recursively generate the
infinite-dimensional contracted Onsager
algebra:
\begin{align}
 \left[Z_k,\:\bar Z_l\right]&=B_{k+l+1},&
 \left[Z_k,\:B_l\right]&=Z_{k+l},&
 \left[B_k,\:\bar Z_l\right]&=\bar Z_{k+l},
 \notag\\[-2.5mm]\label{A}\\[-2.5mm]\notag
 \left[Z_k,\:Z_l\right]&=0,&
 \left[\bar Z_k,\:\bar Z_l\right]&=0,&
 \left[B_k,\:B_l\right]&=0,
\end{align}
where $k,l\in\Z_+$ and $B_0=0$ is implied.
In general, the algebra (\ref{A}) admits the
infinite set of the commuting charges \cite{nsusy},
\begin{equation}\label{jmn}
 \mathcal J_n^l=\frac 12\sum_{p=1}^l\left(
 \left\{\bar Z_{p-1},\:Z_{l-p}\right\}-B_pB_{l-p}\right)
 +\frac n2B_l\sigma_3,
\end{equation}
which includes the Hamiltonian (\ref{sH}),
${\cal J}^1_n=2{\cal H}_n$.

The operators $Z$, $\bar Z$ in the representation
(\ref{Z}) with the conformal factor (\ref{S2}) corresponding
to the sphere generate the following finite-dimensional
algebra which we call the {\it intrinsic}:
\begin{eqnarray}
& [Z,\:\bar Z]=G-4\beta^2 {\cal S}_\Phi,&\nonumber\\
&[Z,\:G]=-4\beta^2\D,\qquad
[\bar Z,\: G]=4\beta^2\bar\D,\qquad
[Z,\:\bar\D]=G,\qquad
[\bar Z,\:\D]=-G,&\nonumber\\
&[{\cal S}_\Phi,\:Z]=-Z,\qquad
[{\cal S}_\Phi,\:\bar Z]=\bar Z,\qquad
 [{\cal S}_\Phi,\:\D]=-\D,\qquad
 [{\cal S}_\Phi,\:\bar\D]=\bar\D,&\label{fda}\\
 &[Z,\:\D]=[\bar\D,\:\bar Z]=[\D,\:\bar\D]=
 [{\cal S}_\Phi,\: G]=[G,\:\D]=[G,\:\bar\D]=0,&\nonumber
 \end{eqnarray}
where
\begin{equation}
{\cal S}_\Phi=\hat S-\frac{1}{2}\Phi,\qquad
G=B(\bz z)-2\beta^2\Phi.
\label{cals}
\end{equation}
The operator $G$ satisfies the obvious relation
$\iint_{S^2}G\,dv=0$, whereas
the multiplicative operators
$\D$ and $\bar\D$ are,
in correspondence with Eq. (\ref{fda}),
the tensor fields  of helicity $-1$ and $+1$,
respectively, and have the coordinate representation
$\D=\frac 14c_1\bar z\rho(\bz z)$, $\bar\D=\frac 12c_1z$.
The operators $\D$ and $\bar\D$ are
mutually adjoint with respect to the scalar product
(\ref{scal}). The intrinsic algebra (\ref{fda})
and all the
corresponding relations for the hyperbolic plane (\ref{H})
can be reproduced by the formal change $\beta\to i\beta$.

The intrinsic algebra (\ref{fda})
has the following two Casimir
operators:
\begin{align}\label{Casimirs}
 {\cal C}&=G^2+8\beta^2\bar\D\D,&
 {\cal W}&=\bar\D Z+\bar Z \D-\bar\D\D
 -G({\cal S}_\Phi-1).
\end{align}
In the given representation, we have
${\cal C}=\frac 14c_1^2\cdot\I$, while
the operator ${\cal W}$ in the gauge (\ref{gz}) has the form
\begin{align}\label{Wc}
 {\cal W}&=\frac 12c_1\left(z\partial-\bz\bar\partial
 -{\cal S}_\Phi\right).
\end{align}
{}From the coordinate representation (\ref{Wc}),
one can conclude that the operators $\D$ and $\bar \D$
represent the components of the
Killing vector associated with the axial
symmetry, and the Casimir operator ${\cal W}$ is
proportional
to the generator of this symmetry. It is interesting to note
that in terms of the algebra (\ref{fda}),
the symmetry generator itself
has the form $J={\cal W}/{\sqrt{\cal C}}$, i.e.	it defines
``spin'' of representations of the algebra (see below),
and the eigenvalue (\ref{sphi}) of the
operator ${\cal S}_\Phi$ has a sense
of ``effective helicity" shifted
by the quantized magnetic flux.
Therefore, in the case of the odd values of the
magnetic flux, we have here an example of the
``boson-fermion" transmutation.
We shall return to the discussion of this aspect
of the system below.

Rescaling appropriately the generators, it is always
possible to reduce the constant $4\beta^2$ to
$\epsilon=\pm 1$, where $`+$'
corresponds to the sphere, while $`-$' does to the
hyperbolic plane. Then, one can redefine the generators of
the algebra (\ref{fda}) as
\begin{align}\label{E}
 E_-&=Z-\frac 12\mathcal D,&
 E_+&=\bar Z-\frac 12\bar{\mathcal D},&
 E_0&=\epsilon{\cal S}_\Phi,\\\label{T}
 T_-&=\mathcal D,&
 T_+&=\bar{\mathcal D},&
 T_0&=\epsilon G.
\end{align}
{}From (\ref{fda}) it follows that the operators (\ref{T})
span the Abelian algebra of translations, $\t(3)$,
while the operators (\ref{E}) together with (\ref{T})
obey the commutation relations:
\begin{gather*}
[E_+,\,E_-]=\epsilon E_0,\hskip 1cm
[E_0,\,E_\pm]=\pm E_\pm,\hskip 1cm
[E_0,T_\pm]=\pm T_\pm,\\
[E_\pm,\,T_0]=\pm T_\pm,\hskip 1cm
[E_\pm,\,T_\mp]=\mp\epsilon T_0,\hskip 1cm
[E_\pm,\,T_\pm]=[E_0,\,T_0]=0.
\end{gather*}
Therefore, one can conclude that
in the case of the sphere
the intrinsic algebra (\ref{fda}) is
the algebra of the 3D Euclidean group of motions,
$\mathfrak{iso}(3)=\t(3)\oplus_s\mathfrak{so}(3)$,
while in the case of the hyperbolic plane
it is the Poincar\'e algebra,
$\mathfrak{iso}(2,1)=\t(3)\oplus_s\mathfrak{so}(2,1)$.
In this context, the
first Casimir operator in (\ref{Casimirs}) is the
``squared mass'' operator,
${\cal C}=T_0^2+2\epsilon T_-T_+$, while the second
one is the Pauli-Lubanski ``pseudo-scalar'',
${\cal W}=\epsilon T_0(1-E_0) + T_+ E_-+E_+T_-$.

In terms of the algebra (\ref{fda}), the generators of
the contracted Onsager algebra (\ref{A}) read as
\begin{gather}
 Z_k=(-4\beta^2)^k(Z+k\D),\hskip 2cm
 \bar Z_k=(-4\beta^2)^k(\bar Z+k\bar \D),
 \notag\\[-2.5mm]\label{gens}\\[-2.5mm]\notag
 B_k=(-4\beta^2)^{k-1}(kG-4\beta^2{\cal S}_\Phi),
\end{gather}
Hence, we have obtained the nontrivial realization of the
contracted Onsager algebra
in the sense that though the infinite set of
generators (\ref{gens})
is represented linearly in terms of
the
finite number of the
intrinsic algebra generators,
nevertheless, $Z_k$, $\bar Z_k$ and $B_l$ do not
turn into zero for $k\ge 1$, $l\ge 2$.
In this context one notes that
in the
systems with nonlinear holomorphic supersymmetry
investigated earlier \cite{d1,d2}, only the non-contracted
Onsager algebra case was realized nontrivially \cite{nsusy}.

The operators $Z_k$, $\bar Z_k$
with $k\ge 1$ also obey the contracted
Dolan-Grady relations. Hence, one can try to use them to
define a ``new'' physical system with the Hamiltonian of the
same form (\ref{sH}). Since the operators $\D$ and $\bar\D$
are the tensors of helicity $-1$ and $+1$, the generating
elements
$Z_k$, $\bar Z_k$ of the ``new'' system can be treated as
a
deformation of the original $U(1)$-connection.
As a result,
the ``new'' system constructed in such a way
turns out to be equivalent to the ``old'' one
up to a redefinition of the parameters $c_0$, $c_1$
and
$\beta$.

Among the central charges (\ref{jmn}), besides the
Hamiltonian ${\cal H}_n=\frac 12\mathcal J_n^1$, the
operator $\mathcal J_n^2$ is the only independent integral
of motion, which can be represented in the form
\begin{equation}\label{j2}
 \mathcal J_n^2={}-4\beta^2\left(2\mathcal J_n^1+
 {\cal W}\right)
 -\frac{1}{2}(4\beta^2)^2{\cal S}_\Phi\left
 ({\cal S}_\Phi+n\sigma_3\right)
 -\frac 12{\cal C}.
\end{equation}
Hence, in the representation under consideration, the
Hamiltonian ${\cal H}_n$ and the Casimir operator ${\cal W}$
form the set of independent central charges of the
nonlinear superalgebra \cite{nsusy}, which are the
differential operators.
We also note that the obvious integrals of motion, $\hat S$
and $\sigma_3$, of the supersymmetric system
can be transformed into the central charges
${\cal T}_1=2\hat S+n\sigma_3$ and
${\cal T}_2=\hat S(\hat S+n\sigma_3)$,
and in terms of them, the
second structure in the integrals (\ref{j2}) is represented
linearly,
$$
{\cal S}_\Phi
 ({\cal S}_\Phi+n\sigma_3)=
 \frac{1}{4}\Phi^2-\frac{1}{2}\Phi{\cal T}_1+{\cal T}_2.
$$
All the commuting charges $\mathcal J_n^l$ (\ref{jmn})
with $l\geq 3$ are not independent since they can be
represented as:
$$
 \mathcal J_n^l=(-4\beta^2)^{l-2}
 \left(4\beta^2\left(l-2\right)l{\cal J}_n^1+
 C_2^l{\cal J}_n^2
 +\frac 12 (4\beta^2)^2C_2^{l-1}
{\cal S}_\Phi \left({\cal S}_\Phi+n\sigma_3\right)
 -\frac 12C_3^l\cdot {\cal C}\right),
$$
where $C^l_n=\frac{l!}{(l-n)!n!}$, $l,n\in\N$.

Thus, from the above algebraic analysis it follows
that the operators ${\cal W}$, ${\cal T}_1$
and ${\cal T}_2$ exhaust the list of independent
nontrivial central charges of the $n$-HSUSY
system described by the Hamiltonian ${\cal H}_n$
and by the supercharges (\ref{sQ}).
The given systems on
the sphere and hyperbolic plane provide the first examples
of the nontrivial realization of the contracted Onsager
algebra in contrast with the cases of
the plane \cite{d2} and 1D systems with $n$-HSUSY
\cite{d1}, where the realization of the algebra
turns out to be trivial \cite{nsusy}.

In conclusion of  this section,
let us return to the
integral of motion  $J={\cal W}/{\sqrt{\cal C}}$
constructed from the Casimir operators of the intrinsic
algebra.  Its coordinate form is
\begin{equation}
J=z\partial-\bz\bar\partial
 -{\cal S}_\Phi.
 \label{totj}
 \end{equation}
Since the $J$ is associated with
the axial symmetry of the system, it has a sense
of the
total angular momentum operator.
Then, the shifted helicity  ${\cal S}_\Phi$
(\ref{cals})
can be treated as the ``effective spin", controlling the
phase of the quantum wave functions
that they acquire under the
$2\pi$-rotation of the system.
In the spherical  case,
its eigenvalue $s_\Phi$
(\ref{sphi})
is half-integer when
the quantized magnetic flux $\Phi$
takes an odd value,
i.e. we have here some kind of bose-fermi transmutation.
On the other hand, in the case of the hyperbolic plane
the magnetic flux is not quantized,
the $s_\Phi$ can take arbitrary values,
and one can say that the quantum states of the system
are anyonic-like.
However, such an interpretation has a weak point:
the operator $J$
was obtained by us from the intrinsic algebra,
whose $\mathfrak{iso}(3)$ (or $\mathfrak{iso}(2,1)$ in
the hyperbolic case) generators $E_\pm$, $T_\pm$,
$T_0$
are not integrals of motion, and, so,
do not correspond to any symmetry
of the system. Therefore, it seems that nothing prohibits
to shift  (\ref{totj}), say,  by a constant term
$f(c_0,c_1)$ such that $f(0,0)=0$.
Below, however, we shall
give an additional
argument in favour of treating
(\ref{totj}) as a total angular momentum operator.

%%%%%%%%%%%%%%%%%%%%%%%%%%%%
\section{Algebraic approach to analysis of symmetries in an
external field}

In the previous section we have established that the
integral of motion corresponding to the axial symmetry of
the system is the Casimir operator of the intrinsic algebra
generated by the covariant derivatives
(\ref{fda}). As a consequence,
the operator $J={\cal W}/{\cal C}$ automatically
commutes with the Hamiltonian constructed in terms
of covariant derivatives.
In contrast with the generic case (\ref{B}),
the constant ($c_1=0$) magnetic field does not break the
isometry of the sphere (hyperbolic plane). Therefore, the
system defined
by the Hamiltonian (\ref{sH}) with $B=const$
has to have three integrals
of motion corresponding to the isometry of the background
space.
Here we analyse the symmetries of the system
in the constant external magnetic field in the context
of the corresponding intrinsic covariant algebra.
This will help us to
understand better the question of fixing the form of the
total angular momentum operator (\ref{totj}).
Moreover,
our analysis will result in formulation of the general
method (beyond the
context of supersymmetry)
of obtaining the explicitly covariant form of the
integrals of motion.

Using the representation (\ref{zH})
of the
``bosonic'' Hamiltonian in the gauge (\ref{gz})
with $c_1=0$, one can
find
the coordinate form of the three integrals
of motion,
\begin{align}
 J_+&={}-\beta^2z^2\partial-\bar\partial
 +\left(2s-\frac\Phi 2\right)\beta^2z,
 \notag\\\label{Jc}
 J_0\,&=z\partial-\bz\bar\partial-s+\frac\Phi 2,
 \\\notag
 J_-&=\partial+\beta^2\bz^2\bar\partial
 -\frac\Phi 2\beta^2\,\bz,
\end{align}
where we assume that these operators act on
the tensors of
helicity $s$. Here, the constant term in the operator $J_0$
is completely fixed by the condition that the operators
(\ref{Jc})
realize a unitary representation of the
non-Abelian algebra
$\mathfrak{su}(2)$ being the isometry
algebra of the sphere. The
modification corresponding to the
hyperbolic plane is
reproduced, as always,
by the formal change $\beta\to i\beta$,
and in this case
the integrals span the algebra
$\mathfrak{su}(1,1)$. The detailed discussion
of different aspects of the
systems on the sphere and hyperbolic
plane in the constant magnetic field can be found in Refs.
\cite{dunne,iengo,hall}.

The integrals (\ref{Jc}) are given in the non-covariant
form. Nevertheless, we know that they have to be
the scalar
operators since we consider the spectral problem on the
tensors of the definite rank, $s$, and the symmetry,
generated by the integrals, corresponds to the degeneracy of
the spectrum. The coordinate form (\ref{Jc}) and the
information about the type of the operators is enough to
properly define their behaviour on the (co)tangent bundle,
but the covariance with respect to the corresponding
$U(1)$-bundle is still vague. Below,
modifying appropriately
the algebraic
method used in the previous section,
we develop a general
method of finding the explicitly covariant
form of the integrals.

Let us consider the algebra generated by the operators $Z$
and $\bar Z$ in the case of the constant magnetic field,
\begin{align}\label{isom}
 [Z,\:\bar Z]&=-4\beta^2{\cal S}_\Phi,&
 [{\cal S}_\Phi,\:Z]&=-Z,&
 [{\cal S}_\Phi,\:\bar Z]&=\bar Z.
\end{align}
Here the operator ${\cal S}_\Phi$
is formally the same as for the
algebra (\ref{fda}), and
$\beta$ is assumed to be pure imaginary in
the case of the hyperbolic plane.
One can note that this algebra is
exactly the isometry algebra of the background space.
 As we will see later,
this is a general (local) property of the spaces of constant
sectional curvature when external fields are absent or
constant.

Unlike the case of the inhomogeneous magnetic field
(\ref{B}), in the algebra (\ref{isom}) generated by the
covariant derivatives only one additional element appears.
Obviously, we cannot construct the integrals (\ref{Jc}) only
in
terms of the operators $Z$, $\bar Z$ and ${\cal S}_\Phi$.
In
the inhomogeneous case the components of the Killing vector
corresponding to the axial symmetry were
the necessary
constituents of the covariant construction of the integral
$\cal W$ (\ref{Casimirs}). For the homogeneous magnetic
field one needs to use the three Killing vectors.
The form of their components in the coordinate system we are
working in is the following:
\begin{align*}
 V_0&=\frac 12\rho(\bz z)\bz,&\bar V_0&=z,&
 V_1&=\frac 12\rho(\bz z)\beta^2\bz^2,&
 \bar V_1&=\beta^2z^2,&
 V_2&=\frac 12\rho(\bz z),&\bar V_2&=1.
\end{align*}
Under the change of the coordinates
the $V_i$ ($\bar V_i$), $i=0,1,2$, are
transformed as the tensors
of helicity $-1$ ($+1$). We will treat
$V_i$  and $\bar V_i$
as multiplication operators for the given quantum
system. In this context, the operators $V_i$ and $\bar V_i$
are mutually adjoint with respect to the scalar product
(\ref{scal}). Together with the covariant derivatives, $V_i$
and $\bar V_i$ generate the following algebra:
\begin{gather}
 [Z,\:\bar V_0]=\phantom{-}G_0,\hskip 7mm
 [\bar Z,\: V_0]=-G_0,\hskip 7mm
 [Z,\: G_0]=-4\beta^2V_0,\hskip 7mm
 [\bar Z,\:G_0]=\phantom{-}4\beta^2\bar V_0,
 \notag\\\notag
 [Z,\:\bar V_1]=\phantom{-}G_1,\hskip 7mm
 [\bar Z,\: V_1]=-\bar G_1,\hskip 7mm
 [Z,\: G_1]=\phantom{-}4\beta^2V_2,\hskip 7mm
 [\bar Z,\:G_1]=\phantom{-}4\beta^2\bar V_1,
 \notag\\[-3mm]\label{alg2}\\[-3mm]\notag
 [Z,\:\bar V_2]=-\bar G_1,\hskip 7mm
  [\bar Z,\: V_2]=\phantom{-}G_1,\hskip 7mm
 [Z,\: \bar G_1]=-4\beta^2V_1,\hskip 7mm
 [\bar Z,\:\bar G_1]=-4\beta^2\bar V_2,
 \notag\\\notag
 [{\cal S}_\Phi,\:V_i]=-V_i,\hskip 5.5mm
 [{\cal S}_\Phi,\:\bar V_i]=\bar V_i,
  \notag\\\notag
 [Z,\:V_i]=[\bar Z,\:\bar V_i]=[{\cal S}_\Phi,\:G_\alpha]
 =[{\cal S}_\Phi,\:\bar G_\alpha]=0,
\end{gather}
where $i=0,1,2$, $\alpha=0,1$, and
in the chosen coordinate system,
$\bar G_0=G_0$, $G_1$
and
$\bar G_1$ are the
scalar multiplicative operators of the form
\begin{align*}
 G_0&=2\sqrt{\rho(\bz z)}-1,&
 G_1&=2\beta^2z\sqrt{\rho(\bz z)},&
 \bar G_1&=2\beta^2\bz\sqrt{\rho(\bz z)}.
\end{align*}
Since $V_i$, $\bar V_i$,
$G_\alpha$ and $\bar G_\alpha$ are
the multiplicative operators,
they form an Abelian subalgebra. Moreover, from the
commutation relations (\ref{alg2}) it follows that this
Abelian subalgebra is an ideal. Finally, one can conclude
that the algebra given by the commutation relations
(\ref{isom}), (\ref{alg2}) is
$\mathfrak g=\bigl(\mathfrak{t}_-(3)\oplus\mathfrak{t}_0(3)
\oplus\mathfrak{t}_+(3)\bigl)\oplus_s\mathfrak{isom}$, where
$\mathfrak{t}_-(3)=\mathsf{span}\{V_1,\bar G_1,\bar V_2\}$,
$\mathfrak{t}_0(3)=\mathsf{span}\{V_0, G_0,\bar V_0\}$,
$\mathfrak{t}_+(3)=\mathsf{span}\{V_2, G_1,\bar V_1\}$,
and $\mathfrak{isom}$ is the isometry algebra of the sphere
or of the hyperbolic plane. The Abelian subalgebras
$\mathfrak{t}_\pm(3)$ are mutually conjugate with respect
to the scalar product (\ref{scal}). The algebra
$\mathfrak g$ has seven Casimir operators constructed from
the
generators $\mathfrak{t}_{\pm}(3)$, $\mathfrak{t}_0(3)$.
They are trivial in the given representation and their
particular form is not important for the further
discussion.

The Hamiltonian (\ref{sH}) is composed of the covariant
derivatives, which belong to the algebra $\mathfrak{isom}$.
Therefore, the Casimir operators of the subalgebras
$\mathfrak{t}_{\pm}(3)\oplus_s\mathfrak{isom}$
and
$\mathfrak{t}_0(3)\oplus_s\mathfrak{isom}$, each of which is
isomorphic to $\mathfrak{iso}(3)$ (sphere) or to
$\mathfrak{iso}(2,1)$ (hyperbolic plane),
\begin{align}
 J_+&=\bar ZV_2-\bar V_1Z
 +\left({\cal S}_\Phi-1\right)G_1,
 \notag\\\label{Jcov}
 J_0\,&=\bar V_0Z+\bar ZV_0
 -\left({\cal S}_\Phi-1\right)G_0,
 \\\notag
 J_-&=\bar V_2Z-\bar ZV_1
 +\left({\cal S}_\Phi-1\right)\bar G_1,
\end{align}
commute with the Hamiltonian (\ref{sH}). These scalar
operators are explicitly covariant with respect to the
$U(1)$ and (co)tangent bundles and in the gauge (\ref{gz})
their coordinate form coincide with the operators
(\ref{Jc}). Thus, we have found the covariant form of
the integrals
of motion for the given system (with $B=const$) in the pure
algebraic manner.

It is worth noting that in the given representation the
multiplicative operators obey some additional
relations
of a polynomial form. We will discuss only the
most
interesting of them. For example, one can verify
that the
multiplicative operators satisfy the constraints
\begin{align*}
 2\beta^2V_0-G_1V_1-\bar G_1V_2&=0,&
 (1-G_0)V_2-G_1V_0&=0,&
 (1+G_0)V_1-\bar G_1V_0&=0,
\end{align*}
and their conjugate. These constraints mean that the
Killing vectors are not independent in some sense. The
commutation of these relations with the covariant
derivatives leads to another set of constraints:
\begin{align*}
 G_1&=4\beta^2\left(\bar V_0V_2+\bar V_1V_0\right),&
 \bar G_1&=4\beta^2\left(\bar V_0V_1+\bar V_2V_0\right),&
 G_0&=2\left(\bar V_2V_2-\bar V_1V_1\right).
\end{align*}
Obviously, one can use these relations to exclude the
operators $G_0$, $G_1$ and $\bar G_1$ from the algebra
(\ref{isom}), (\ref{alg2}). Then, one can say that the
operators $Z$, $\bar Z$, $V_i$ and $\bar V_i$ generate some
{\it quadratic} algebra. Therefore, the quantum system in
the homogeneous magnetic field represents
a physical system with
the intrinsic nonlinear algebra.

In the case of an inhomogeneous magnetic field with the
axial symmetry, the components of only one of the Killing
vectors, say, $V_0$ and $\bar V_0$, can be considered as
the proper
multiplicative operators of the corresponding quantum
system. Thus, for the axial magnetic
field of a general form
only the operators $Z$, $\bar Z$, $V_0$ and
$\bar V_0$ can be considered as generating elements of the
algebra, intrinsic to the given quantum system. In general
case, the covariant derivatives $Z$ and $\bar Z$ generate an
infinite-dimensional subalgebra. Nevertheless, the class of
systems with a finite-dimensional generated algebra should
be large enough. Note, that the appearance of such a
finite-dimensional algebra implies that the magnetic field
obeys some conditions (like those appearing
from the Dolan-Grady
relations). The algebraic approach for obtaining
the corresponding integrals of motion can be applied in
these
cases as well.

In addition to the gauge interaction, one can
include into this scheme the interaction with a scalar
potential, which, in general, should be treated as an
independent generating element of the intrinsic algebra. As
a toy example, one can include into the Hamiltonian
(\ref{sH}) the
interaction with the scalar potential of the form
$U=U(G_0,\bar V_0V_0)$. Obviously, such a potential does not
change the intrinsic algebra $\mathfrak{g}$ (\ref{isom}),
(\ref{alg2}), but it breaks the symmetry of the quantum
system to the axial subgroup generated by the operator
$J_0$.

In the spherical case,
when the magnetic field is switched off,
the generators (\ref{Jc}), forming the
$\mathfrak{su}(2)$ algebra, realize the representations with
integer weight only since we start form the wave functions
to be tensor fields, i.e. initially spin of the system is
integer.
When the magnetic field is switched on,
in accordance with  the coordinate form of $J_0$ (\ref{Jc}),
the spin of the quantum
system effectively undergoes the shift by $\Phi/2$.
Hence, if the flux
parameter $\Phi$ is odd, the spin of the system is
half-integer. This phenomenon
is well-known, e.g., in the context
of the charged system immersed into the field of a
magnetic monopole \cite{hooft}.
At the same time, the 2D quantum
system  on the sphere subjected to
the homogeneous magnetic field can
be treated as a 3D system
in the field of magnetic monopole
reduced to the spherical geometry.
In the case of the
hyperbolic plane the real parameter $\Phi$ is not
restricted, and
the generators (\ref{Jc})
forming the noncompact
$\mathfrak{su}(1,1)$ algebra realize
a representation with arbitrary real weight
\cite{tor} defined
by the shifted helicity operator
$S_\Phi$.

The magnetic field of the form (\ref{B}) possesses only the
axial symmetry associated with the Killing vector $V_0$,
$\bar V_0$. It means that the components of the other
Killing vectors cannot be considered as a proper generating
elements any more. Moreover, in the case (\ref{B}) the
multiplicative operators $V_0$, $\bar V_0$ are not
independent generating elements since the intrinsic algebra
(\ref{fda}) is generated merely by the covariant
derivatives. Indeed, the operators $\cal D$ and
$\bar{\cal D}$ in the algebra (\ref{fda})
are proportional, correspondingly, to $V_0$ and $\bar V_0$.
The integral $\cal W$ (\ref{Casimirs})
associated with the axial symmetry,
being the Casimir of the algebra
$\mathfrak t(3)\oplus_s\mathfrak{isom}$,
is similar to the integrals (\ref{Jc}).
In this context, one
can say that the presence of the inhomogeneous
part in the magnetic field (\ref{B})
breaks the algebra $\mathfrak g$ (\ref{isom}), (\ref{alg2})
to its subalgebra
$\mathfrak t_0(3)\oplus_s\mathfrak{isom}$,
and from the three Casimir operators
(\ref{Jc}) only
the operator
$J_0$ survives
in the form of the integral (\ref{totj}).
Such a relation between the cases
with inhomogeneous axially symmetric
magnetic field and homogeneous magnetic field
supports our treatment of the operator
(\ref{totj})
as the 2D total angular momentum operator
of the system (\ref{sH}), (\ref{Z}), (\ref{B}).

The above algebraic approach can be generalized to the case
of curved spaces of higher dimensionality. To argue this,
let
us discuss a quantum system in a $D$-dimensional Riemannian
space with a constant sectional curvature \cite{kn},
the wave
functions of which are covariant tensors. Then the Riemann
tensor has the form
$$
 R_{\mu\nu}{}^{\alpha\beta}=\frac{\cal R}{D(D-1)}
 \left(\delta_\mu^\alpha\delta_\nu^\beta-
 \delta_\mu^\beta\delta_\nu^\alpha\right),
$$
where the constant $\cal R$ is the scalar curvature.
Obviously, such a space is a symmetric one and its isometry
algebra is maximal: $\mathfrak{so}(D+1)$ (${\cal R}>0$),
$\mathfrak{iso}(D)$ (${\cal R}=0$) or
$\mathfrak{so}(D,1)$ (${\cal R}<0$). The covariant
derivatives with the Riemann connection generate the
corresponding isometry algebra. Indeed, their commutator
can be represented in the form
$$
 [\nabla_\mu,\:\nabla_\nu]=\frac 12
 R_{\mu\nu}{}^{\alpha\beta}M_{\alpha\beta}=
 \frac{\cal R}{D(D-1)}M_{\mu\nu},
$$
where the operators $M_{\mu\nu}$ are given in a matrix
representation defined by the
fields which the covariant
derivatives act on. For example, the action of the operators
on a contravariant vector field reads as
$M_{\mu\nu}\circ V^\alpha=(\delta^\alpha_\mu
g_{\nu\gamma}-\delta^\alpha_\nu
g_{\mu\gamma})V^\gamma$, while for a spinorial field the
corresponding matrix is proportional to
$[\gamma_\mu,\:\gamma_\nu]$, where $\gamma_\mu$ are
the
$D$-dimensional $\gamma$-matrices. The generalization to
the
fields of arbitrary type is straightforward. It is not
difficult
to verify that the covariant derivatives and the operators
$M_{\mu\nu}$ form the algebra isomorphic to the isometry
algebra of the background space, i.e. the covariant
derivatives are the
generating elements of the isometry algebra.
The number of Killing vector fields
$V_a=V_a^\mu(x)\partial_\mu$ forming the isometry algebra is
equal to $D(D+1)/2$. Let us suppose that the interaction
with an external gauge field or with a scalar potential
breaks this algebra to a subalgebra spanned by a subset of
Killing vectors $V_i$, $i=1,\ldots, l<D(D+1)/2$. Treating
the components $V_i^\mu$ and the scalar potential (if it is
present) as multiplicative operators of the
given quantum system, one can consider them together with
the covariant derivatives as generating elements of the
intrinsic algebra of the system. The corresponding
Hamiltonian is supposed to be composed of the covariant
derivatives and of the scalar potential. Evidently, if the
resulting algebra is finite-dimensional, then the covariant
form of the corresponding integrals of motion can be found
in closed terms.

%%%%%%%%%%%%%%
\section{Discussion and outlook}

Let us summarize briefly the obtained results and discuss
some open problems that deserve further attention.

We have discussed the application of the general algebraic
scheme of the nonlinear holomorphic supersymmetry to the
quantum mechanical systems with nontrivial 2D Riemann
geometry in the presence of external magnetic field. The
analysis of Dolan-Grady relations (\ref{dgr}), underlying
the construction \cite{nsusy}, shows that
\begin{itemize}
\item
The nonlinear holomorphic supersymmetry can be realized on
Riemann surfaces with constant curvature only.
\end{itemize}
Moreover, for non-vanishing curvature the $n$-HSUSY can be
realized solely in the case of the contracted Dolan-Grady
relations (\ref{cdgr}). We have investigated the cases of
the sphere and Lobachevski plane, which together with the
zero curvature case of the plane analysed by us earlier
\cite{d2}, exhaust the simply connected Riemann surfaces. We
demonstrated that for the $n$-HSUSY system on the sphere and
hyperbolic plane a part of the spectrum can be found
algebraically by reducing the corresponding 2D Hamiltonian
to one-dimensional families of the QES Hamiltonians
described by the $\mathfrak{sl}(2,\R)$ scheme.

We have found that the one-dimensional QES potential
corresponding to the sphere admits a discrete group of
transformations of the parameters of the potential. The 1D
potential corresponding to the case of hyperbolic plane
admits a similar discrete group of transformations if
some parameter is restricted to be integer. In particular,
this means that
\begin{itemize}
\item
 The discrete transformations (\ref{trans})--(\ref{duali})
 partially relate distinct sectors of the spectrum of the
 same 2D system,
 or the spectra of
 different 2D systems.
\end{itemize}
Due to the second relation in (\ref{duali}), the discrete
transformations are a kind of duality transformations.
We argued that, in general, other QES $\mathfrak{sl}(2,\R)$
potentials also can  admit such a symmetry group after
discretization of some parameters. This was illustrated by
the example of the QES potential (\ref{x6}). We are going to
investigate in detail the revealed discrete symmetry of QES
potentials elsewhere.

The analysis of the algebraic content of the model on the
sphere and hyperbolic plane shows that all the observables
of the systems can be represented in terms of the generators
of the $\mathfrak{iso}(3)$ (sphere) or
$\mathfrak{iso}(2,1)$ (hyperbolic plane) intrinsic
algebras. All the higher generators of the contracted
Onsager algebra associated with the relations
(\ref{cdgr}) are linearly dependent
(see Eq. (\ref{gens})) but
non-vanishing. Hence,
\begin{itemize}
\item
 The supersymmetric 2D systems on the sphere and hyperbolic
 plane realize the nontrivial representations of the
 contracted Onsager algebra.
\end{itemize}
Using the representation (\ref{gens}), we have found that
besides the Hamiltonian, the infinite set of commuting
charges (\ref{jmn}) contains, in fact, only one non-trivial
independent integral of motion, which is a differential
operator being, simultaneously, the Pauli-Lubanski
``pseudo-scalar'' $\cal W$ (\ref{Casimirs})
of the corresponding $\mathfrak{iso}(3)$ or
$\mathfrak{iso}(2,1)$ algebra.
It plays  the role of the 2D total angular momentum
(\ref{totj}),
and its noncoordinate part
has a sense of the effective spin of the system.
The value of the effective spin ${\cal S}_\Phi$
(\ref{cals}),
being a helicity shifted by the magnetic flux
parameter, takes integer or half-integer values on the
sphere, and corresponds to the
anyonic-like quantum states
on the hyperbolic plane.

The algebraic analysis allowed us to find
the covariant form of the
integral of motion associated with
the axial symmetry. Following the same pattern we
have found the
covariant form of all the integrals in the case of the
homogeneous
magnetic field.
On the basis of these explicit examples, we
have
proposed
\begin{itemize}
 \item
 A general algebraic method to find the  covariant form of
 integrals of motion of a quantum system in a curved space
 in the presence of an external gauge field.
\end{itemize}
The integrals obtained in this way have transparent
commutation properties. We hope that this approach will be
helpful for a wide class of quantum systems with symmetry.

Nowadays, a considerable attention is
paid to a generalization
of the 2D systems on Riemann surfaces with a
constant magnetic field \cite{iengo,li,hall,dunne}
to the case of noncommutative surfaces
\cite{nc1,nc2,nc3,nc4}.
Such quantum systems are exactly solvable in the both
cases of commutative and noncommutative spaces.
The supersymmetric systems
considered in this paper can be treated as a generalization
of the constant magnetic field case to the case
of the nontrivial field ($B\neq const$) accompanied by
the transition from exactly solvable
systems to quasi-exactly solvable ones. Therefore, the
interesting problem is to generalize the 2D realizations of
the nonlinear holomorphic supersymmetry to the cases of the
noncommutative plane, sphere and hyperbolic plane.
Such a generalization will be presented
elsewhere \cite{noncom}.

\vskip 0.5cm
{\bf Acknowledgements}
\vskip 5mm

One of us (S. K.) thanks J. Gamboa for useful
communications.
The work was supported by the grants 1010073 and 3000006
from FONDECYT (Chile) and by DICYT (USACH).

%%%%%%%%%%%%%%%%%%%%%%%%%%%%%%%%%%%

\end{document}